\documentclass[12pt]{article}
\usepackage{authblk}
\usepackage[bookmarksnumbered, colorlinks, plainpages]{hyperref}
\usepackage{amsmath, amsthm, amscd, amsfonts, amssymb, graphicx, color, booktabs, cite}
\numberwithin{equation}{section}
\definecolor{email}{rgb}{0.00,0.00,0.84}
\begin{document}
\setcounter{page}{1}

\title{\large \bf 12th Workshop on the CKM Unitarity Triangle\\ Santiago de Compostela, 18-22 September 2023 \\ \vspace{0.3cm}
\LARGE Lattice outlook on $B\to\rho\ell\bar{\nu}$ and $B\to K^\star \ell \ell$ \normalsize}

\author[1,2]{Luka Leskovec${}^{\star}$}
\author[3]{Stefan Meinel}
\author[4]{Marcus Petschlies}
\author[5]{John Negele}
\author[6]{Srijit Paul}
\author[5]{Andrew Pochinsky}
\author[7]{Gumaro Rendon}

\affil[1]{Faculty of Mathematics and Physics, University of Ljubljana, Jadranska 19, 1000}
\affil[2]{Jozef Stefan Institute, Jamova 39, 1000 Ljubljana, Slovenia}
\affil[3]{Department of Physics, University of Arizona, Tucson, AZ 85721, USA}
\affil[4]{Helmholtz-Institut f\"ur Strahlen- und Kernphysik, Rheinische Friedrich-Wilhelms-Universit\"at Bonn, Nußallee 14-16, 53115 Bonn, Germany}
\affil[5]{Center for Theoretical Physics, Massachusetts Institute of Technology, Cambridge, MA 02139, USA}
\affil[6]{ Maryland Center for Theoretical Physics, University of Maryland, College Park, USA}
\affil[7]{Fujitsu Research, Santa Clara, California, USA}

\maketitle

\begin{abstract}

Lattice Quantum Chromodynamics (QCD) has significantly contributed to our understanding of the CKM matrix through precise determinations of hadronic matrix elements.  With advancements in theoretical methodologies and computational resources, investigations can now extend to processes involving QCD-unstable hadrons such as the $\rho$ and $K^\star(892)$. These resonances play vital roles in processes such as weak decays of $B$ mesons, opening new avenues for exploration. Finite-volume lattice QCD techniques involving complex computational methods are used to determine the transition amplitudes. Here, we present preliminary results for  $B\to\rho\ell\bar{\nu}$.
\end{abstract} \maketitle

\section{Introduction}

\noindent  Lattice Quantum Chromodynamics (QCD) has been pivotal in furthering our understanding of the CKM matrix. Through collaborative endeavors and targeted experimental campaigns on a global scale, the community determined many parameters of the Standard Model, including $|V_{ud}|$, $|V_{us}|$, $|V_{cd}|$, and $|V_{cs}|$ with high precision \cite{FlavourLatticeAveragingGroupFLAG:2021npn}. These endeavors, however, have primarily focused on processes where the initial and final hadron states are stable under QCD. Developments in theoretical approaches, as well as the availability of computational resources, allow us to extend calculations to processes that involve QCD-unstable hadrons, i.e., hadronic resonances, beyond the narrow-width approximation. Two of the most interesting of such processes are the charged-current decay of a $B$ meson to a $\rho$ resonance and a pair of lepton and its anti-neutrino, $B\to\rho\ell\bar{\nu}$, and the rare decay of a $B$ meson to a $K^\star$ resonance and a pair of lepton and anti-lepton: $B\to K^\star \ell \ell$. Previous lattice calculations of these processes were performed in the narrow-width approximation \cite{UKQCD:1995uhp,Flynn:2008zr,Bowler:2004zb,Bowler:1993rz,Bowler:1994cq,Burford:1995fc,DelDebbio:1997kr,Bernard:1993yt,Abada:1995fa, Becirevic:2006nm,Horgan:2013hoa,Horgan:2013pva}. \\

The $\rho$ and the $K^\star(892)$ are similar in that they are both elastic vector resonances ($J^P=1^-$) that decay to a single two-hadron channel. The $\rho$ has isospin $I=1$ and decays to a pair of pions in $p$-wave, while the $K^\star(892)$ has isospin $I=1/2$ and decays to a pion and a kaon in $p$-wave. In fact, in lattice QCD, studying resonances means studying the scattering in a given channel, where the resonances then appear as poles of the scattering amplitude $T$. From that point of view, we are then interested in two types of scattering - the first is $\pi\pi$ scattering with isospin $I=1$, and the second is $K\pi$ scattering with isospin $I=1/2$. The former process, $\pi\pi$ scattering in $I=1$, has only a single relevant partial wave $\ell$ contribution below $1$ GeV, the $p$-wave. The $s$- and $d$-wave do not contribute due to Bose symmetry, and the $f$-wave contribution is tiny due to the threshold behavior requiring $k^{2\ell + 1}$, where $k$ is the $\pi\pi$ scattering momentum. The latter process, $K\pi$ scattering in $I=1/2$, has two relevant contributions, the $s$-wave, where the $K_0^\star(700)$ resonance resides, and the $p$-wave with the $K^\star(892)$ resonance \cite{Leskovec:2012gb}. The $d$-wave resonates above the $1$ GeV limit at approximately $1.4$ GeV, where the $K_2^\star(1430)$ is.\\

The scattering amplitudes for the two-hadron systems then enter the hadronic matrix element of the $B\to\rho\ell\bar{\nu}$ and $B\to K^\star \ell \ell$ through the rescattering process. The hadronic matrix elements involving the two-hadron final state can be decomposed as
\begin{align}
\label{eq:H}
\langle h_1 h_2(P,\ell) | J^\mu | B(p) \rangle = \sum_i K{^\mu}_i F_i(E^\star, q^2) \frac{T(E^\star)}{k^\ell},
\end{align}
where $h_1$ and $h_2$ are the two hadrons in the final state with total four-momentum $P$ and in partial wave $\ell$, $J^\mu$ is the weak current, and the initial-state is a $B$-meson with four-momentum $p$. The kinematic prefactors of the Lorentz decomposition $K_i^\mu$ differ between different types of currents inserted and dictate the complexity of the analysis. We choose to express the hadronic matrix element in terms of functions of the two kinematic variables, $E^\star$ - the two-hadron energy in the center-of-momentum frame (CMF), and $q^2=(p-P)^2$. To take care of the two-hadron rescattering threshold behavior, the scattering amplitude is divided by $k^\ell$, where $k$ is the two-hadron scattering momentum. In Eq.~\eqref{eq:H}, $F_i(E^\star, q^2)$ are the generalized transition form factors. They are smooth functions of $E^\star$ and $q^2$ in the studied region, while $T(E^\star)$ - the scattering amplitude of the two-hadron system, contains all the two-hadron analytic structures: the resonance poles, the scattering branch cuts, and, if necessary, any left-hand-cut physics. The quantities of interest to both experiment and theory are the transition amplitudes $F_i(E^\star, q^2) \frac{T(E^\star)}{k^\ell}$, and they are present in both the matrix elements determined on the lattice and the partial branching fractions observed in the experiment. Here, we briefly describe how such a transition amplitude can be obtained and what results can be expected from lattice QCD calculations for the example of the vector-current contribution to the $B\to\rho\ell\bar{\nu}$ process. In this case, the decomposition becomes
\begin{align}
    \langle \pi \pi, \epsilon(P,s) | V^\mu | B(p) \rangle = \frac{2 i V(E^\star,q^2)}{m_B + 2m_\pi} \varepsilon^{\mu\nu\alpha\beta} \epsilon(P,s)^*_\nu P_\alpha p_\beta,
\end{align}
where
\begin{align}
V(E^\star,q^2) = F(E^\star, q^2) \frac{T(E^\star)}{k},
\end{align}
and $\epsilon(P,s)$ is the polarization vector of the $\ell=1$ final state labeled by a spin index $s$. In the following, we present preliminary results on a single gauge-field ensemble with $N_f = 2 + 1$ clover-Wilson fermions whose quark masses correspond to $m_\pi \approx 320$ MeV. The lattice spacing is approximately $a = 0.114$ fm. For the $b$-quark, we use an anisotropic action \cite{Chen:2000ej,El-Khadra:1996wdx}, in which we tune the quark mass and anisotropy parameters to match the $B_s$ meson rest and kinetic mass.

\section{Lattice QCD and the finite-volume tool}

Lattice-QCD calculations are performed in a finite volume, where the fields obey periodic boundary conditions in space, with cubic symmetry or its subgroups. As a direct result of the finite volume, the spectrum of states in a lattice QCD calculation is no longer continuous but rather discrete. Even more, the QCD states themselves get affected by the finite volume, yielding two major effects that need to be accounted for in the analysis of lattice-QCD data.\\

The first effect is the energy shifts, as first pointed out by L\"uscher and further developed by many authors. In short, the finite-volume energies appear as poles of the energy-dependent two-point correlation function, and the location of the poles is related to the infinite-volume scattering matrix $T$ through the quantization condition \cite{Luscher:1990ux,Rummukainen:1995vs,Kim:2005gf,Briceno:2014oea,Briceno:2017max,Woss:2020cmp}:
\begin{align}
    \det \left[ F^{-1}(E^\star) + T(E^{\star}) \right]\vert_{E^\star = E_n^\star} = 0,
\end{align}
where $F^{-1}$ is a linear combination of the L\"uscher Zeta functions and accounts for the finite-volume symmetries, $T$ is the scattering matrix for the given channel, and $E_n^\star$ is the center-of-momentum frame discrete set of energies as determined from the lattice calculation. In this manner, the scattering amplitude can be determined, and further details on the $\rho$ and $K^\star$ resonances discussed here can be found in Refs.~\cite{Alexandrou:2017mpi} and \cite{Rendon:2020rtw}.\\

The second effect is the normalization of the states, first pointed out by Lellouch and L\"uscher: as the states appear as poles, they will naturally be classified by their position (related to the energy) and their residue. The latter is related to the normalization of the finite-volume state \cite{Lellouch:2000pv,Lin:2001ek,Briceno:2014uqa,Briceno:2015csa,Briceno:2021xlc}:
\begin{align}
    \label{eq:norm}
    |E_n^\star \rangle_L = \sqrt{R} |h_1 h_2(E^\star=E_n^\star)\rangle_\infty,
\end{align}
where $R$ is the residue of the pole and can be determined as
\begin{align}
    R=\lim_{E^\star \to E_n^\star} \frac{E^\star - E^\star_n}{ F^{-1}(E^\star) + T(E^{\star})}.
\end{align}
In Eq.~\eqref{eq:norm}, $ |h_1 h_2(E^\star=E_n^\star)\rangle_\infty$ is the infinite-volume two-particle state at its invariant mass being equal to the lattice energy. While not shown explicitly here, the states need to be projected to definite irreducible representations of the relevant subgroup of the cubic group.\\

\begin{figure}[htb!]
    \centering
    \includegraphics[width=0.8\textwidth]{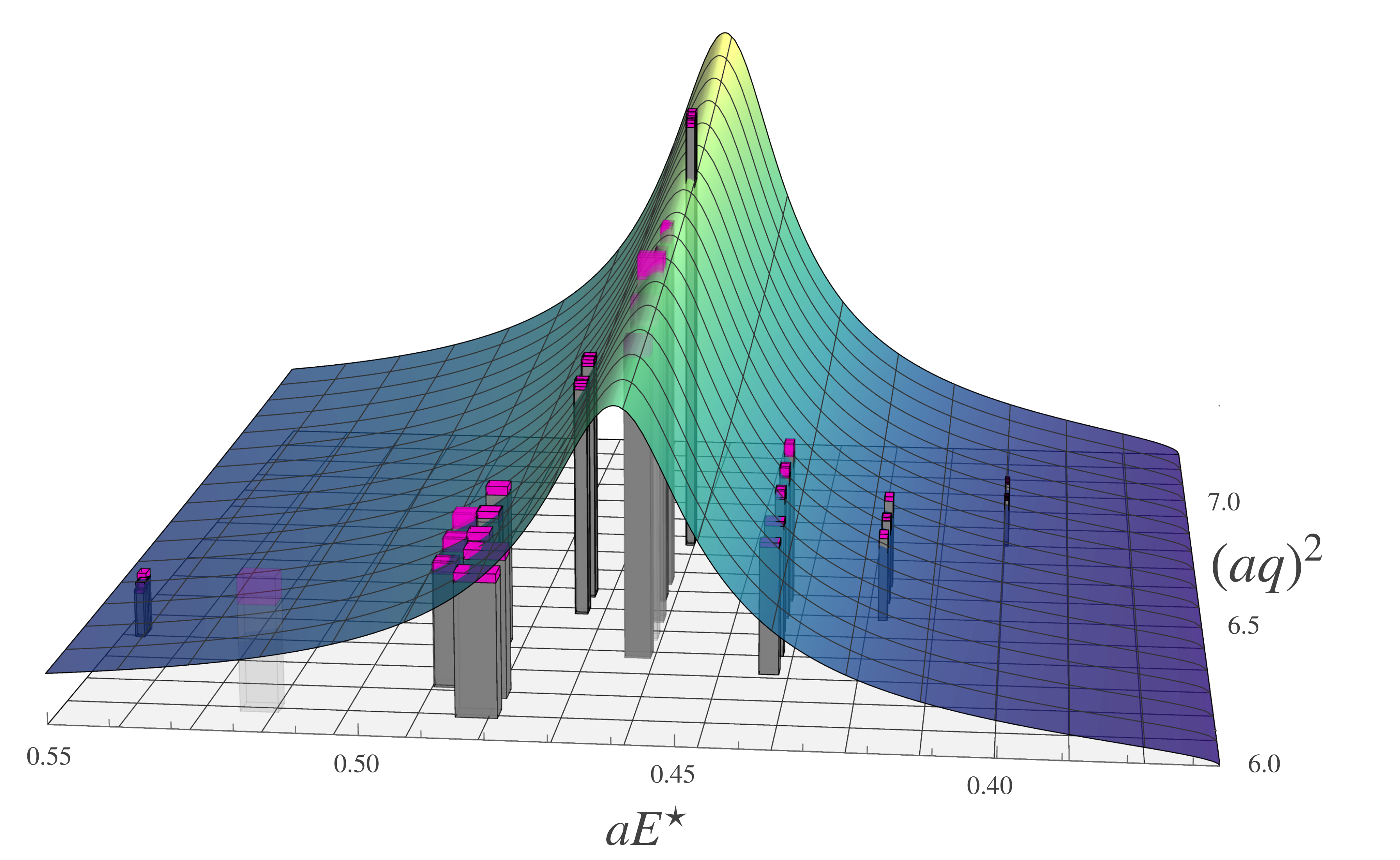}
    \caption{Our preliminary results for the transition amplitude $V(E^\star,q^2)$, using the central parametrization ``F3N1M1\_TBWII''. Here, the bars show the lattice results at the available kinematic points, with magenta sections indicating the statistial uncertainties.}
    \label{fig:3d}
\end{figure}

The energies $E_n$, and hence the scattering amplitudes and pole residue can be extracted from the Euclidean-time dependence of $\pi\pi$ two-point correlation functions. The matrix elements of the weak current are determined from three-point correlation functions. These finite-volume matrix elements are related to the infinite-volume matrix elements through \cite{Leskovec:2022ubd}
\begin{align}
   \langle \pi \pi, \epsilon(P,s) | V^\mu | B(p) \rangle_L = \sqrt{R} \langle \pi \pi, \epsilon(P,s) | V^\mu | B(p) \rangle_{\infty}.
\end{align}
We perform global fits of the matrix elements at all available kinematic points using suitable parametrizations for $T$ and $F$. For $T$, we use the Breit-Wigner models ${\rm BWI}$ and ${\rm BWII}$ of Ref.~\cite{Alexandrou:2017mpi}, while for $F$ we use two families of parametrizations that generalize the $z$ expansion (in which the variable $q^2$ is mapped to the new variable $z$ that takes on values in the unit disk) \cite{Boyd:1994tt,Bourrely:2008za,Alexandrou:2018jbt}:
\begin{itemize}
 \item[{F1)}] Combined order $K$: 
 \begin{align}
    F(q^2,E^\star) = \frac{1}{1 - \frac{q^2}{m_P^2}} \sum_{n+m\leq K} A_{nm} z^n \mathcal{S}^m,
 \end{align}
  
 \item[{F3)}] Order $N$ in $z$, order $M$ is $\mathcal{S}$:
 \begin{align}
    F(q^2,E^\star) = \frac{1}{1 - \frac{q^2}{m_P^2}} \sum_{n=0}^{N} \sum_{m=0}^{M} A_{nm} z^n \mathcal{S}^m.
 \end{align} 
\end{itemize}
Above, $\mathcal{S}=\frac{(E^\star - 2m_\pi)^2}{4m_\pi^2}$, and $m_P=m_{B^*}$ for the vector form factor. Altogether, we consider $10$ parametrizations that satisfy the following conditions: $ \frac {\chi^2}{\rm dof} < 1.5$, and all parameters are  resolved from zero in at least one of the scattering-amplitude models, ${\rm BWI}$, or ${\rm BWII}$. As our central parametrization we choose ``F3N1M1\_TBWII'', i.e. F3 with $N=1$ and $M=1$ and ${\rm BWII}$, as our central parametrization. A plot of the transition amplitude $V(E^\star,q^2)$, using the central parametrization is shown in Fig.~\ref{fig:3d} in the region of $q^2$ and $E^\star$ where lattice data is available.

\begin{figure}[htb!]
    \centering
    \includegraphics[width=0.75\textwidth]{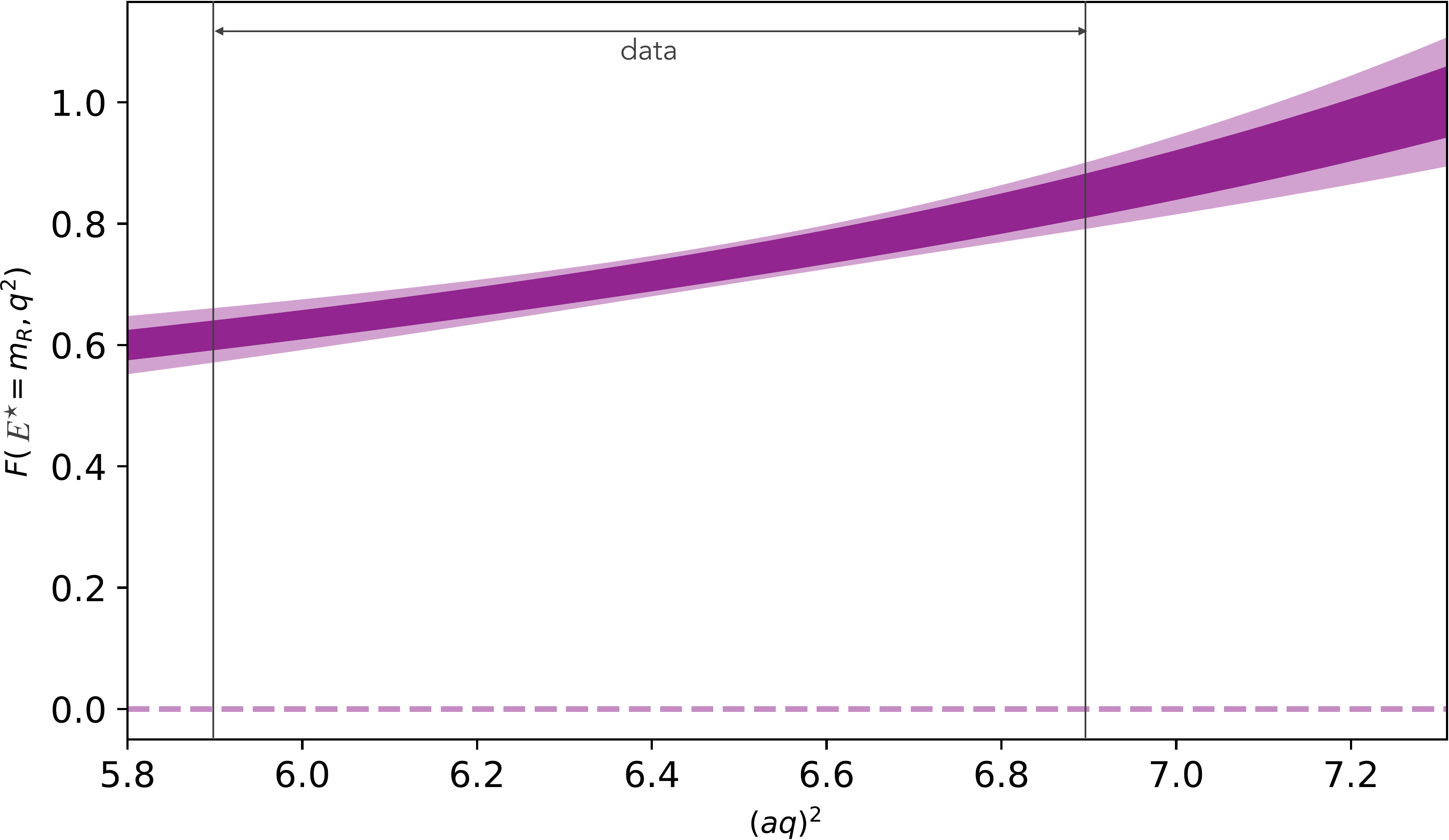}
    \caption{Our preliminary results for the function $F(E^\star, q^2)$, with $E^\star$ set to the resonance mass $m_R$ (note that this is not the $B\to\rho$ resonance form factor). The inner shaded band shows the statistical uncertainty and the outer band shows the combined statistical and parametrization uncertainty. The region where lattice data is available is indicated. }
    \label{fig:combine}
\end{figure}

A plot of the function $F(E^\star, q^2)$, with $E^\star$ set to the resonance mass $m_R$, is shown in Fig.~\ref{fig:combine} (note that this is not the $B\to\rho$ resonance form factor). To gauge the degree of parametrization dependence we compute the root-mean-square deviation among the different parametrizations at each $q^2$. We find approximately $4\%$ statistical uncertainty and approximately $5\%$ parametrization uncertainty at highest $q^2$ for which we have data. The resonance form factor can be obtained through an analytic continuation of the transition amplitude $V(E^\star,q^2)$ to complex $E^\star$ and extracting the residue of the $\rho$ pole \cite{Briceno:2021xlc}.

\section{Summary} \label{sec:Summary}

We have presented preliminary results for the vector form factor of the process $B\to\pi\pi\ell\bar{\nu}$ with the $I=1$, $\ell=1$ $\pi\pi$ final state. For the ensemble used here, with a pion mass of $320$ MeV, we have achieved an approximately $6\%$ statistical and parameterization uncertainty in the high-$q^2$ region. The analysis of the axial form factors is more involved, because the three different form factors, $A_0$, $A_1$ and $A_2$, all appear in a single matrix element. This analysis is ongoing. Calculations on three additional gauge ensembles at different spacings and pion masses are underway to enable chiral and continuum extrapolations. In summary, we have demonstrated that analyses of heavy-light $1\to 2$ transition form factors are feasible and reasonable precision can be obtained. This encourages further investigation in this process as well as other processes such as the $B\to K\pi \ell \ell$ as well as $B\to D\pi \ell \bar{\nu}$ \cite{Gustafson:2023lrz}.

\section*{Acknowledgments}

We thank Kostas Orginos, Balint Joó, Robert Edwards, and their collaborators for providing the gauge-field configurations. Computations for this work were carried out in part on (1) facilities of the USQCD Collaboration, which are funded by the Office of Science of the U.S.~Department of Energy, (2) facilities of the Leibniz Supercomputing Centre, which is funded by the Gauss Centre for Supercomputing,  (3) facilities at the National Energy Research Scientific Computing Center, a DOE Office of Science User Facility supported by the Office of Science of the U.S.~Department of Energy under Contract No.~DE-AC02-05CH1123, (4) facilities of the Extreme Science and Engineering Discovery Environment (XSEDE), which was supported by National Science Foundation grant number ACI-1548562, and (5) the Oak Ridge Leadership Computing Facility, which is a DOE Office of Science User Facility supported under Contract DE-AC05-00OR22725. L.L.~acknowledges the project (J1-3034) was financially supported by the Slovenian Research Agency. S.M.~is supported by the U.S. Department of Energy, Office of Science, Office of High Energy Physics under Award Number D{E-S}{C0}009913. J.N.~and A.P.~acknowledge support by the U.S. Department of Energy, Office of Science, Office of Nuclear Physics under grants DE-SC-0011090 and DE-SC0018121 respectively. A.P.~acknowledges support by the “Fundamental nuclear physics at the exascale and beyond” under grant DE-SC0023116.

\bibliographystyle{utphys-noitalics}
\bibliography{pos}

\end{document}